\documentclass[12pt]{article}
\usepackage{epsfig}
\makeatletter
\def\alt{\mathrel{\mathpalette\gl@align<}}
\def\agt{\mathrel{\mathpalette\gl@align>}}
\def\gl@align#1#2{\lower.6ex\vbox{\baselineskip\z@skip\lineskip\z@
\ialign{$\m@th#1\hfil##\hfil$\crcr#2\crcr\sim\crcr}}}
\makeatother

\begin{document}
\begin{flushright}
{\tt hep-ph/0302176}\\
OSU-HEP-03-4 \\
February, 2003 \\
\end{flushright}
\vspace*{2cm}
\begin{center}
{\baselineskip 25pt
\large{\bf
Unification of Gauge, Higgs and Matter in Extra Dimensions
}}

\vspace{1cm}

{\large
Ilia Gogoladze\footnote
{On a leave of absence from: Andronikashvili Institute of Physics, GAS, 380077, Tbilisi, Georgia.\\
email: {\tt ilia@hep.phy.okstate.edu}}, %and
Yukihiro Mimura\footnote
{email: {\tt mimura@hep.phy.okstate.edu}}
and
S. Nandi\footnote
{email: {\tt shaown@okstate.edu}}
}
\vspace{.5cm}

{\small {\it Physics Department, Oklahoma State University, \\
             Stillwater, OK 74078}}

\vspace{.5cm}

\vspace{1.5cm}
{\bf Abstract}
\end{center}

We consider the unification of gauge, Higgs as well as the matter fields
in a 6D $N\!\!=2$ supersymmetric $SU(8)$ gauge theory.
The gauge symmetry $SU(8)$ is broken down to $SU(4) \times SU(2)_L \times SU(2)_R \times U(1)^2$
in 4D through $T^2/Z_6$ orbifold compactification,
and the theory is reduced to 4D $N=1$ supersymmetric Pati-Salam model.
The electroweak Higgs fields as well as the third family of fermions are
unified in the 6D $N=2$ gauge multiplet.
The 6D bulk gauge interaction provides both
gauge and Yukawa interactions for the third family
predicting $\alpha_1 = \alpha_2 = \alpha_3 = \alpha_t = \alpha_b = \alpha_\tau$
at the unification scale, in good agreement with experiment.
Incorporation of the first and second family as well as other orbifolds are also briefly discussed.

\thispagestyle{empty}

\bigskip
\newpage

\addtocounter{page}{-1}

\section{Introduction}
\baselineskip 20pt

Recent topics of the theories in higher dimensions give us a lot of interesting
phenomenological pictures.
One of the most attractive motivations of extension of dimensions
is that the variety of particles in Nature
can be understood by means of a geometrical language.
For example, gauge fields with the extra dimensional components
behave as scalar fields in 4 dimension.
Since masses of the gauge bosons are prohibited by gauge invariance,
the scalar field originated from the gauge bosons can be a good candidate
for the low energy Higgs fields which break electroweak symmetry.
That leads to the idea of the gauge-Higgs unification in the
higher dimensional theories \cite{Manton:1979kb,Hosotani:1983xw,Hall:2001tn}.

We consider that the extra dimensions are compactified in an orbifold.
% in order to make chiral theories in 4D,
%since 5D fermions include both chirality in 4D language.
%If we consider the higher dimensional theories on the orbifolding
In an orbifold space, % such as $S^1/Z_2$,
we can impose transformation properties to the fields %at the folding places,
and the symmetries can be broken %through these boundary conditions
\cite{Hosotani:1983xw,Scherk:1978ta}.
%
%Recent realization of the phenomenological models in higher dimensions
%makes us encourage to revisit
%this idea \cite{Krasnikov:dt,Hall:2001zb,Burdman:2002se,Haba:2002vc,Gogoladze:2003bb}.
%
Recently, a great deal of works has been done on
%many people have a great interest of
the gauge symmetry breaking using the orbifold compactification,
and these lead to many attractive features of the unified gauge theories
in higher dimensions \cite{Kawamura:1999nj,Hebecker:2001jb}.
Using the orbifold transformation properties, we can project out unwanted fields
such as colored Higgs triplets in the grand unified theories \cite{Kawamura:1999nj}.
In such a progress of the higher dimensional unified theories,
many people have revisited the idea of
gauge-Higgs unification \cite{Krasnikov:dt,Hall:2001zb,Burdman:2002se,Haba:2002vc,Gogoladze:2003bb}.
In the higher dimensional supersymmetric theories,
the gauge multiplet contains both vector multiplet and
chiral supermultiplets in 4D.
Assigning the different transformation property between vector multiplet
and chiral supermultiplets,
we can make vector multiplet massless but chiral supermultiplets heavy,
which means the supersymmetry is broken.
If we break gauge symmetry through boundary condition simultaneously,
a part of the chiral supermultiplets can have a zero mode
which remains massless in the low energy.
Then, we can identify such a supermultiplet with the low energy Higgs field.
%This is the main idea of the gauge-Higgs unification which we consider in this paper.
This idea is featured in the 6D $N=2$ supersymmetric theories \cite{Hall:2001zb},
and more recently in the 5D $N=1$ supersymmetric theories \cite{Burdman:2002se,Haba:2002vc,Gogoladze:2003bb}.
The Ref.\cite{Burdman:2002se} emphasizes an interesting possibility
that the gauge and Yukawa coupling constants have the same origin.
Since the Yukawa interactions arise from the gauge interaction
in the 5D lagrangian,
those two coupling constants are ``unified" in the 5D theory.
%This is a very interesting feature of the their gauge-Higgs unified scenario.
In the Ref.\cite{Burdman:2002se}, the authors considered the theory of $SU(3)_w$ and $SU(6)$
as an example of this scenario.
%but in their models, there are many unwanted fields in the matter hypermultiplets.
%In order to make the unwanted fields heavy, they need many brane-localized fields
%for each generation.
%The brane fields are actually needed to cancel the gauge anomaly in the 4D theory
%which arises from the zero modes of bulk hypermultiplets,
%and that means it is not easy to understand the anomaly free structure in their models.
%In Ref.\cite{Haba:2002vc},
%The authors consider
The gauge-Higgs
unification in larger gauge group such as $E_6$, $E_7$ and $E_8$
has also been studied \cite{Haba:2002vc}
In Ref.\cite{Gogoladze:2003bb}, the authors consider the
gauge-Higgs unification in the $SU(4)_w$ and $SO(12)$, and suggest
that the left-right symmetric breaking of the gauge symmetry gives
an economical realization to the representation of the quarks and
leptons.

Another interesting possibility of unified model in higher dimensions
is that quarks and leptons can be unified in the gauge multiplet.
Three families of quarks and leptons can be contained in the
adjoint representations in large gauge groups, such as $E_7$ and $E_8$.
The matters in the adjoint representation are always vector-like,
but one can project out the vector-like partner
by $Z_3$ transformation properties \cite{Babu:2002ti}.
This encourages us to consider that the gauge and matter (quarks and leptons)
unification in
higher dimensional models.
In other context,
we can consider the origin of the three families as
chiral superfields in gauge multiplet \cite{Watari:2002tf}
since the gauge multiplet in the 6D $N=2$ supersymmetry
contains three $N=1$ chiral superfields in 4D.

In this paper,
we consider the possibility of unifying gauge, matter and Higgs fields
in one supersymmetric gauge multiplet in higher dimensions,
as well as the unification of the gauge and Yukawa couplings.
As a simple example,
we construct a 6D $N=2$ supersymmetric $SU(8)$ unified model
where such an unification is achieved.
The gauge symmetry $SU(8)$ is broken down to
%Pati-Salam symmetry,
$SU(4) \times SU(2)_L \times SU(2)_R \times U(1)^2$
in 4D through $T^2/Z_6$ orbifold compactification,
and the theory is reduced to 4D $N=1$ supersymmetric Pati-Salam model \cite{Pati:1974yy}.
The electroweak Higgs fields and standard model fermions for the 3rd family
can be unified with the gauge bosons in the 6D gauge multiplets.
The 6D bulk gauge interaction produces
Yukawa interactions, which give masses to the quarks and leptons
by Higgs mechanism,
and gives gauge-Yukawa unification.
%We will exhibit the numerical results of the renormalization group evolution
%of the Yukawa couplings.
The numerical agreement of this gauge-Yukawa unification prediction
for all the gauge and the third family Yukawa couplings is good.

Our paper is organized as follows:
In section 2, we construct our supersymmetric $SU(8)$ model in 6D
with gauge, Higgs and matter unification and show
how orbifold compactification leads to Pati-Salam model in 4D.
%In section 3, we discuss the quark and lepton mass matrices
%in our model and various other possibility.
%In section 3, we will show the
Numerical results
for gauge and Yukawa coupling unification are shown in section 3.
Section 4 contains our conclusions and discussions.

%\newpage

\section{The Model}

%In this section, we construct the 6D $N=2$ supersymmetric $SU(8)$ gauge theory

We consider the 6D gauge theory with $N=2$ supersymmetry.
The two extra dimensions are compactified by the orbifold $T^2/Z_n$ \cite{Li:2001dt}.
The $N=2$ supersymmetry in 6D corresponds to $N=4$ supersymmetry
in 4D, thus only the gauge multiplet can be introduced in the bulk.
In terms of 4D $N=1$ language, the gauge multiplet
contains vector multiplet $V(A_\mu,\lambda)$ and three chiral multiplets
$\Sigma$, $\Phi$ and $\Phi^c$ in the adjoint representation
of the gauge group.
The fifth and sixth components of the gauge fields, $A_5$ and $A_6$,
are contained in the lowest component of $\Sigma$, {\it i.e.}
$\Sigma|_{\theta=\bar \theta=0} = (A_6 + i A_5)/\sqrt2$.

The bulk action, written in the 4D $N=1$ language and in the Wess-Zumino gauge,
is given by \cite{Arkani-Hamed:2001tb}
\begin{eqnarray}
S &=& \int d^6x \left\{ {\rm Tr} \left[ \int d^2 \theta \left(
\frac1{4kg^2} W^\alpha W_\alpha + \frac1{kg^2} \left( \Phi^c \partial \Phi
- \sqrt2 \Sigma [\Phi,\Phi^c] \right) \right) + h.c. \right] \right.\nonumber \\
&& + \int d^4 \theta \frac1{kg^2} {\rm Tr} \left[ (\frac1{\sqrt2} \partial^\dagger +\Sigma^\dagger)
e^{-2V} (-\frac1{\sqrt2} \partial + \Sigma) e^{2V} + \frac14 \partial^\dagger e^{-2V} \partial e^{2V} \right]
\nonumber \\
&& + \left. \int d^4 \theta \frac1{kg^2} {\rm Tr} \left[\Phi^\dagger e^{-2V} \Phi e^{2V}
+ \Phi^{c \dagger} e^{-2V} \Phi^c e^{2V} \right] \right\},
\label{6D_action}
\end{eqnarray}
where $k$ is the normalization of the group generator, ${\rm Tr} T^a T^b = k \delta^{ab}$ (we take $k=1/2$),
$\partial$ is defined as $\partial = \partial_5 - i\partial_6$,
and $W_\alpha$ is defined as $W_\alpha = - \frac18 \bar D^2 (e^{-2V} D_\alpha e^{2V})$.
The 6D gauge transformations are
\begin{eqnarray}
e^{2V} &\rightarrow& e^{\Lambda} e^{2V} e^{\Lambda^\dagger}, \quad
\Sigma \rightarrow e^{\Lambda} (\Sigma - \frac1{\sqrt2} \partial ) e^{-\Lambda},
\label{gauge-transform}\\
\Phi   &\rightarrow& e^{\Lambda} \Phi e^{-\Lambda}, \quad
\Phi^c \rightarrow e^{\Lambda} \Phi^c e^{-\Lambda}.
\label{gauge-transform2}
\end{eqnarray}

The $T^2/Z_n$ orbifold is constructed by identifying the complex coordinate $z$ of the extra dimensions
under $Z_n : z \rightarrow \omega z$, where $\omega^n =1 $.
The number $n$ is restricted to be $n=2,3,4,6$.
We can impose the transformation property of the gauge multiplet as
\begin{eqnarray}
V(x^\mu,\omega z, \bar \omega \bar z) &=& R \ V(x^\mu,z,\bar z) R^{-1},
\label{transformation1} \\
\Sigma(x^\mu,\omega z, \bar \omega \bar z) &=& \bar \omega \ R \ \Sigma(x^\mu,z,\bar z) R^{-1},
\label{transformation2} \\
\Phi(x^\mu,\omega z, \bar \omega \bar z) &=& \omega^{l} R \ \Phi(x^\mu,z,\bar z) R^{-1},
\label{transformation3}\\
\Phi^c(x^\mu,\omega z, \bar \omega \bar z) &=& \omega^{m} R \ \Phi^c(x^\mu,z,\bar z) R^{-1},
\label{transformation4}
\end{eqnarray}
where $R$ is a unitary matrix and satisfies that $R^n$ is the identity matrix.
Non-trivial $R$ breaks the gauge symmetry.
Because of the lagrangian invariance in Eq.(\ref{6D_action})
under the transformations (\ref{transformation1}-\ref{transformation4}),
we have a relation $l+m \equiv 1 $ (mod $n$).
In the case $n>2$, this transformation property breaks $N=4$ supersymmetry down to $N=1$ in 4D.

Now we consider the $SU(8)$ gauge symmetry and $T^2/Z_6$ orbifold.
We choose the $8\times 8$ matrix $R$ as
\begin{equation}
R = {\rm diag} (1,1,1,1,\omega^5,\omega^5,\omega^2,\omega^2).
\label{unitary R}
\end{equation}
With this choice, $SU(8)$ breaks down to $SU(4) \times SU(2)_L \times
SU(2)_R \times U(1)^2$, and the theory is reduced to 4D $N=1$
supersymmetric Pati-Salam model with two extra $U(1)$ symmetry.
The $SU(8)$ adjoint representation $\mathbf{63}$ is decomposed under
$SU(4) \times SU(2)_L \times SU(2)_R \times U(1)^2$ representations as
\begin{equation}
\mathbf{63} = \left(
\begin{array}{ccc}
\mathbf{(15,1,1)}_{0,0} & \mathbf{(4,2,1)}_{2,0} & \mathbf{(4,1,2)}_{2,4} \\
\mathbf{(\bar 4,2,1)}_{-2,0} & \mathbf{(1,3,1)}_{0,0} & \mathbf{(1,2,2)}_{0,4} \\
\mathbf{(\bar 4,1,2)}_{-2,-4} & \mathbf{(1,2,2)}_{0,-4} & \mathbf{(1,1,3)}_{0,0}
\end{array}
\right) + \mathbf{(1,1,1)}_{0,0} + \mathbf{(1,1,1)}_{0,0},
\end{equation}
where the subscripts denote the charges under the $U(1)_1 \times U(1)_2$ symmetry.
The $Z_6$ transformation property for these respective decomposed representations of the vector multiplet
$V$ and chiral multiplet $\Sigma$ is
\begin{equation}
V : \left(
\begin{array}{ccc}
1 & \omega & \omega^4 \\
\omega^5 & 1 & \omega^3 \\
\omega^2 & \omega^3 & 1
\end{array}
\right) + (1) + (1),
\quad
\Sigma: \left(
\begin{array}{ccc}
\omega^5 & 1 & \omega^3 \\
\omega^4 & \omega^5 & \omega^2 \\
\omega & \omega^2 & \omega^5
\end{array}
\right) + (\omega^5) + (\omega^5).
\label{trans-law}
\end{equation}
We see from (\ref{trans-law}) that the gauge fields
for the unbroken symmetry group
have massless modes in $V$,
and $\mathbf{(4,2,1)}_{2,0}$ component in $\Sigma$ has massless mode.
Choosing $l=4$ and $m=3$,
we obtain the transformation property for $\Phi$ and $\Phi^c$ as
\begin{equation}
\Phi : \left(
\begin{array}{ccc}
\omega^4 & \omega^5 & \omega^2 \\
\omega^3 & \omega^4 & \omega \\
1 & \omega & \omega^4
\end{array}
\right) + (\omega^4) + (\omega^4),
\quad
\Phi^c: \left(
\begin{array}{ccc}
\omega^3 & \omega^4 & \omega \\
\omega^2 & \omega^3 & 1 \\
\omega^5 & 1 & \omega^3
\end{array}
\right) + (\omega^3) + (\omega^3),
\end{equation}
and thus $\mathbf{(\bar 4,1,2)}_{-2,-4}$ in $\Phi$,
$\mathbf{(1,2,2)}_{0,4}$ and $\mathbf{(1,2,2)}_{0,-4}$ in $\Phi^c$ have
massless modes in 4D.
We identify $\mathbf{(4,2,1)}$ and
$\mathbf{(\bar 4, 1, 2)}$ as left- and right-handed quarks and leptons respectively
in one family,
and two bidoublets $\mathbf{(1,2,2)}$ as Higgs fields which break electroweak symmetry.
Then Higgs fields and one family of matter fields are unified
with the gauge fields in the 6D $N=2$ gauge multiplet.
We denote those fields as
\begin{equation}
\Psi_L : \mathbf{(4,2,1)}_{2,0},
\quad
\Psi_R : \mathbf{(\bar 4,1,2)}_{-2,-4}
\quad
H_1 : \mathbf{(1,2,2)}_{0,4},
\quad
H_2 : \mathbf{(1,2,2)}_{0,-4}.
\label{massless-particle}
\end{equation}

We note briefly the other choices of $l$, $m$ and unitary matrix
$R$. We may have three solutions of $l+m \equiv 1$ (mod $6$),
$(l,m) \equiv (0,1),(5,2),(4,3)$ without loss of generality.
(Exchanging of $l$ and $m$ gives identical sets of zero-modes.)
Only the choice of $(l,m)= (4,3)$ leads to one family of matter as the
zero-modes. With the different choice of $R$ from
Eq.(\ref{unitary R}), contents of zero-modes will be different. We have chosen
the unitary matrix $R$ so that both left- and right-handed
matters are chiral, and at that time, we get two Higgs bidoublets as
zero-modes. We have also chosen the left-handed matter is in the
chiral multiplet $\Sigma$.

Since the three $N=1$ chiral multiplets $\Sigma$, $\Phi$ and $\Phi^c$ are in the gauge multiplets,
those fields have gauge interactions with each other in 6D.
This interaction term from Eq.(\ref{6D_action}) is
\begin{equation}
S = \int d^6 x \left[\int d^2 \theta \frac1{kg^2} {\rm Tr} \left(- \sqrt2 \Sigma [\Phi,\Phi^c]\right)
+ h.c. \right],
\label{trilinear term}
\end{equation}
includes Yukawa interaction
\begin{equation}
S = \int d^6 x \int d^2 \theta \ y_6 \Psi_L H_1 \Psi_R + h.c.
\end{equation}
Taking into account the normalization factor of $\Psi_L$, $\Psi_R$ and $H_1$ in the kinetic term,
we find that the six dimensional Yukawa coupling is equal to six dimensional gauge coupling,
$y_6 = g_6$.
The corresponding four dimensional couplings are derived as the coordinates of extra dimensions
are integrated out in the action.
In the ideal situation, the four dimensional Yukawa and gauge coupling can be the same dimensionless
number.
The ideal situation is the following:
1) The brane-localized gauge and Yukawa interactions can be negligible.
2) The zero modes of fermions $\Psi_L$ and $\Psi_R$ are not localized
at different points on the orbifold.
3) The four dimensional fields are not largely mixed with other brane-localized fields.

As for the first situation, the 4D gauge couplings can receive
contributions from brane-localized gauge kinetic terms, such as
$\delta(z) F_{\mu\nu}^2$. However, such contributions can be
expected negligible if the volume of extra dimensions is large.
Since the left-handed fermions $\Psi_L$ is obtained from $\Sigma$
which includes the gauge fields with fifth and sixth coordinates,
the $\Psi_L$ transforms non-linearly under gauge transformation in
Eq.(\ref{gauge-transform}). Thus, we cannot write the
brane-localized Yukawa interaction. If we consider Wilson-line
operator, we may write a gauge invariant 4D interaction, but again
the contribution can be negligible if the volume of the extra
dimensions is large. Next, we consider second ideal situation. If
the zero modes of fermions are localized at different points, the
4D Yukawa couplings become proportional to the overlap of the zero
modes wave functions. In general, vacuum expectation values of the
singlet in the three chiral multiplets gives bulk masses of the
fermions, and then the zero modes might be localized around the
3-brane. But if the vacuum expectation values are smaller than
$1/R$ ($R$ is a radius of the extra dimensions), the contribution
can be negligible. The situation 3) is just the usual things of
the four dimensional model building. If the fields mix with some
other fields and the mass eigenvalues are the linear combination of
the mixed fields, the coupling for the mass eigenstates is the
original couplings multiplied with the mixing angle. Thus, the
mixing angle is small, the Yukawa coupling is almost same as the
gauge couplings.

We identify the one family originating from gauge multiplet with the 3rd family,
and 1st and 2nd family are brane-localized fields at 3-brane fixed point.
Then the Yukawa couplings for 3rd family, $y_t$, $y_b$ and $y_\tau$,
are unified to the gauge couplings at grand unified scale.
We will see the numerical studies in Section 4.
The Yukawa couplings of 1st and 2nd family
are naturally smaller than the 3rd family couplings,
since their values are suppressed by volume factor of the extra dimensions.
%We will see the quark and lepton mass matrices in Section 3.

Since we project out the vector-like partners,
the remaining fermion components
Eq.(\ref{massless-particle})
give gauge anomaly for the extra two $U(1)$ symmetry.
Both two linear combinations of $U(1)$ symmetry have gauge anomaly,
and Green-Schwarz mechanism \cite{Green:sg} can cancel out only one linear combination.
Thus we have to introduce other brane fields to cancel the anomaly.
For example,
if we introduce brane fields with appropriate $U(1)$ charge such as
\begin{equation}
\Psi_L^\prime : \mathbf{(4,2,1)}_{-2,0}, \quad \Psi_R^\prime : \mathbf{(\bar 4,1,2)}_{2,4}
\end{equation}
all gauge anomaly is cancelled out. This can be an origin of other
families. Since one linear combination of $U(1)$ can be anomalous
by using Green-Schwarz mechanism, we can make hierarchical
structure in the quark and lepton mass matrices by
Froggatt-Nielsen like mechanism \cite{Froggatt:1978nt}.

We have made a choice the left-handed matter is in the chiral multiplet $\Sigma$
and right-handed one is in the $\Phi$.
Since the gauge transformations of $\Sigma$ and $\Phi$
in Eqs.(\ref{gauge-transform}-\ref{gauge-transform2})
are different,
brane-localized 4D lagrangian is not left-right symmetric.
The right-handed quark and lepton and Higgs fields
are in the chiral multiplets $\Phi$ and $\Phi^c$
and their gauge transformations are linear,
and thus we can introduce the brane-localized Yukawa coupling terms
which give the CKM mixing angles.
We can also introduce the brane-localized right-handed neutrino mass terms
with the fields $\mathbf{(4,1,2)}$ and/or $\mathbf{(\bar 4,1,2)}$
in the chiral multiplet $\Phi^c$,
and the neutrino masses become small through the seesaw mechanism.
The vacuum expectation values of $\mathbf{(4,1,2)}$ and/or $\mathbf{(\bar 4,1,2)}$
give right-handed Majorana neutrino
masses and also break the Pati-Salam gauge symmetry down to
the standard model gauge symmetry.
We may also use the brane Higgs field $\mathbf{(10,1,3)}$
to give the right-handed neutrinos masses
and to break the Pati-Salam gauge symmetry.

In the Pati-Salam model,
we need at least two Higgs multiplets, otherwise
the down-type quark mass matrix is proportional to up-type quark mass matrix.
In our model, we have two Higgs bidoublets.
Requiring the Goergi-Jarskog relation \cite{Georgi:1979df} for 2nd family,
we need another Higgs field such as $\mathbf{(15,2,2)}$.
For the gauge coupling unification,
only one linear combination of Higgs bidoublet fields should remain at
low energy.
The mixing parameter can be controlled by using the extra $U(1)$ symmetries.

We comment that we can consider three family and Higgs unification
in $T^2/Z^3$ orbifold
if we take $l=m=2$ in Eqs.(\ref{transformation3}-\ref{transformation4}) and
\begin{equation}
R = {\rm diag} (1,1,1,1,\omega^2,\omega^2,\omega,\omega).
\end{equation}
In this case, the origin of the number of family is the number of
chiral multiplets in the $N=4$ gauge multiplet.
%This possibility with respect to the number of family
%has been suggested in Ref.\cite{Watari:2002tf}.
In our model building,
the Yukawa couplings arise from gauge interaction in Eq.(\ref{trilinear term}).
It is easy to see that the mass matrices for the quarks
and leptons are anti-symmetric,
and then one of the family is massless and other two families have same mass.
We can modify this results by adding the brane-localized interactions,
but it is difficult
to split masses of 2nd and 3rd families while preserving
gauge and Yukawa coupling unification.
Of course, we have to care about the $U(1)$ gauge anomaly
and we have to introduce other fields to cancel the anomaly.

%\section{Quark and Lepton Mass Matrices}

%As we noted,
%we consider the case that the one family originated from gauge multiplet is
%identified to 3rd family
%and 1st and 2nd family are brane-localized fields at 3-brane fixed point.

\section{Numerical Results of Gauge and Yukawa Unification}

As was mentioned in the previous section this model can
realize gauge-Yukawa unification for third family quarks and
lepton\footnote{The numerical calculation of gauge-Yukawa
unification in a 4D model is demonstrated in the Ref.\cite{Chkareuli:1998wi}.}.
For realization of this possibility we
assume that the compactification  scale $(M_{U})$ from 6D to 4D
is  the same scale where  $SU(4) \times SU(2)_L \times SU(2)_R
\times U(1)^2$ gauge symmetry are broken to Standard Model one,
choosing appropriate Higgs superfields, which are localized in 4D.
So below $M_{U}$ scale we have usual MSSM particle content with
gauge-Yukawa  unification condition for particles from third
family
\begin{equation}
\alpha_1=\alpha_2=\alpha_3=\alpha_t=\alpha_b=\alpha_{\tau}
\end{equation}
where $\alpha_1$, $\alpha_2$ and $\alpha_3$ corresponds
hypercharge (with proper normalization), weak and strong
interaction couplings. $\alpha_t$, $\alpha_b$, $\alpha_{\tau}$ are the top,
bottom and tau Yukawa coupling respectively. We use the
notation $y^2_{t,b,\tau}/4\pi\equiv\alpha_{t,b,\tau}$. Note that
we neglect brane localized gauge kinetic term.

For numerical calculation of gauge and Yukawa coupling
evolution we are using two-loop RG analysis \cite{br}, with
conversion from $\overline{MS}$ scheme to dimensional reduction
$(\overline{DR})$ one. We include the standard supersymmetric
threshold correction at low energies, taken at a single scale
$M_{SUSY}=M_Z$ \cite{lp}.

Due to a crucial reduction of the number of the fundamental
parameters from the gauge-Yukawa coupling unification, we are lead immediately
to a series of the very distinctive predictions (in absence of any
large supersymmetric threshold corrections). Using the values of the
electroweak parameters $\sin^2\theta_w=0.2311\pm0.0001$ and
$\alpha_{EM}=127.92\pm0.02$ at $M_Z$ scale \cite{PDG},
we can determine the unification scale and unified coupling constant.
%Once we define unification scale
%and unification constant through RG analysis we can predict the
%following quantity
Then, evolving the remaining couplings from the unification scale
to the low energy, we predict
\begin{eqnarray}
\alpha _{3}(M_{Z}) =0.123,~~~m_{t}=178~{\rm GeV},~~~
\frac{m_{b}}{m_{\tau }}(M_{Z}) =1.77,~~~\tan\beta =51. \label{8}
\end{eqnarray}
These are in good agreement with experimental data \cite{PDG}. The
small discrepancy for $\alpha_3$ (world average
value is $\alpha_3=0.117\pm0.002$ \cite{PDG}) can be easily
improved if we consider unification scale threshold of Higgs in
the $\mathbf{(10,1,3)}$ representation, which is brane localized
field. After Pati-Salam symmetry is broken to Standard Model by
$\mathbf{(10,1,3)+(\overline{10},1,3)}$  in the particle spectrum
we have $SU(3)_c$ color sextet $\mathbf{(6+\overline 6)}$ Higgs
field. Arranging for  this fragment mass at the scale $10^{16}$
GeV we can avoid discrepancy of $\alpha_3$ and we have $\alpha_3 =
0.117$. This threshold gives the correction to the top quark mass
to the right direction as unification constant increase from this
correction  and $m_{t}=177~{\rm GeV}$. In Figure 1, unification of
the gauge-Yukawa coupling is demonstrated.
% with correct $\alpha_3$,
%which we adjust with GUT scale threshold.

\begin{figure}
\begin{center}
\epsfig{file=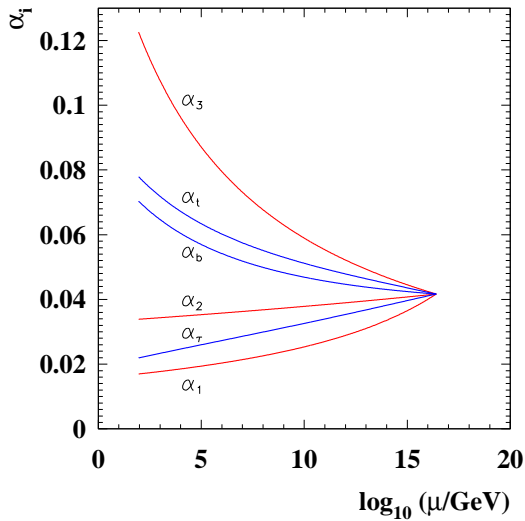,width=8cm}
\end{center}
\caption{ The  unification of gauge ($\alpha_1$, $\alpha_2$,
$\alpha_3$) and Yukawa ($\alpha_t$, $\alpha_b$, $\alpha_{\tau}$)
couplings at SUSY GUT scale (red and blue lines respectively).  }
\label{fig1}
\end{figure}

\section{Conclusion and Discussion}

The idea that the Higgs and/or the matter are unified in
the higher dimensional supersymmetric gauge multiplet
leads us naturally to the world in which the standard gauge group in 4D is unified
in grand unified gauge group in higher dimensions.

We have considered the 6D $N=2$ supersymmetric $SU(8)$ gauge theory
in the orbifold $T^2/Z_6$.
The gauge group $SU(8)$ is broken down
by $Z_6$ transformation property,
and the model is reduced to $N=1$ supersymmetric Pati-Salam model
$SU(4) \times SU(2)_L \times SU(2)_R$
with two extra $U(1)$ symmetries in 4D.
In this model,
two bidoublet Higgs fields and one family of quarks and leptons
are unified in the gauge sector in the 6D $N=2$ gauge supermultiplet.
The vector-like partners of the quarks and leptons
are projected out by $Z_6$ transformation property.
The explicit mass terms of the Higgs fields
are prohibited by the gauge invariance, and the $N=1$ supersymmetry
preserve the masses of the Higgs fields to the electroweak scale.
Furthermore, because of the unification of Higgs and matter in gauge multiplet,
the 4D Yukawa interactions, which we need in order to give masses
to the fermions by Higgs mechanism, arise from gauge interaction
in 6D lagrangian.
This is the most interesting feature of this model.
The one family of quarks and leptons in the gauge multiplet
can be identified to the 3rd family,
and then we meet an attractive possibility
that 3rd family Yukawa couplings can be unified to the gauge coupling
at grand unified scale.
The numerical results of the renormalization group flow of the Yukawa couplings
have a good agreement
with low energy data of the top quark mass and bottom-tau mass ratio
with large $\tan \beta$.

Since one cannot put the additional bulk fields in
6D $N=2$ supersymmetric theory,
1st and 2nd families must be brane-localized fields,
and the Yukawa interactions for 1st and 2nd families
are introduced at the 3-brane.
Thus, the smallness of the masses of the 1st and 2nd families
can be understood as the volume suppression of the Yukawa couplings.
Since both extra $U(1)$ symmetries give gauge anomalies,
we have to introduce additional fields to cancel them.
The additional fields can be identified to another family.
The extra $U(1)$ symmetry can be used to
construct a hierarchical structure of the mass
matrices.

It is interesting that the Yukawa interaction for 3rd family originates
from 6D gauge interaction,
whereas we cannot predict the Yukawa couplings for 1st and 2nd families which are brane fields.
%this model gives us prediction for 3rd family Yukawa couplings.
It is well-known that bottom-tau ratio is successful in the usual $SU(5)$ relation.
%bottom and tau Yukawa couplings are unified in the usual grand unified theories.
In order to realize $m_s/m_\mu$, however, we have to introduce $\mathbf{(15,2,2)}$ Higgs,
which is included in $\mathbf{126}$ or $\mathbf{120}$ in $SO(10)$ language.
Then, a question arises: Why 3rd family doesn't have Yukawa couplings with
such $\mathbf{(15,2,2)}$ Higgs field?
We can answer the question in the context of our gauge-Higgs unification scenario.
That is because $\mathbf{(15,2,2)}$ Higgs is not included in the gauge multiplet
and this Higgs must be brane-localized fields.
The 3rd family Yukawa interactions with $\mathbf{(15,2,2)}$ Higgs are suppressed by volume factor
of extra dimensions.
Thus, we can realize the unification of the bottom-tau Yukawa couplings.

We have used torus $T^2$ as the extra space manifold in this paper.
Zero-modes of matter can be chiral through the $Z_6$ projection.
At the same time, we can have two Higgs bidoublets as zero-modes. One can
consider other manifolds, such as disc $D^2$ or annulus $A^2$. In
that case, we can take $Z_n$ transformation with arbitrary number
of $n$. For example, we can consider $D^2/Z_9$ orbifold and
zero-modes of matter can be chiral through $Z_9$ projection.
In this case, we can have one family and only one bidoublet of the Higgs.
%Then
%only one bidoublet can be zero-mode.

We can also consider other unified gauge groups.
It seems that $SO(12)$ is the smallest rank gauge group
to unify the Higgs and the matter in gauge multiplet.
The $SO(12)$ can be broken down to $SU(5)\times U(1)^2$
by $T^2/Z_6$ orbifold compactification.
We can choose the $Z_6$ projection so that $\mathbf{10}_{2,0}$, $\mathbf{\bar 5}_{-2,2}$,
$\mathbf{5}_{0,2}$ and $\mathbf{\bar 5}_{0,-2}$ have zero modes.
Thus the gauge multiplet contains Higgs fields and one family of matter.
However, 6D gauge interaction in Eq.(\ref{trilinear term})
doesn't include the top quark Yukawa interaction.
Actually, the $U(1)$ symmetry doesn't match in the interaction term $\mathbf{10\cdot 10 \cdot5}$.
Another example of gauge group is $E_7$.
$E_7$ can be broken down to
$E_7 \rightarrow SO(10)\times U(1)^2$
by orbifold compactification.
%If we choose $Z_9$ projection, $\mathbf{16}_{0,-3}$, $\mathbf{16}_{1,1}$ and $\mathbf{10}_{-1,2}$
%can have zero-modes.
%In this case, the 6D gauge interaction contains Yukawa interactions contrary to the $SU(5)$ case.
Since $SU(8)$ is one of the regular maximal subgroup of $E_7$,
our model which we suggest in this paper can be one of the breaking pattern of the $E_7$.
%As for example of another breaking pattern, $E_7$ can be broken down to
%$E_7 \rightarrow SO(10)\times U(1)^2$
%by $T^2/Z_6$ orbifold compactification.

Finally the idea of gauge, Higgs and matter unification is very novel.
The fact that this can be semirealistically achieved in higher dimensional
supersymmetric theories is very intriguing, and leads
to yet another motivation for supersymmetry.

\section*{Acknowledgments}

We thank K.S. Babu, C. Macesanu, J. Lykken and S. Raby for useful
discussions.
%Y.M. acknowledges the warm hospitality and support
%of the KEK Theory Group during his visit there.
This work was supported in part by US DOE Grants \# DE-FG030-98ER-41076
and DE-FG-02-01ER-45684.

\end{document}